\documentclass[%
aip,amsmath,amssymb,reprint]{revtex4-2}

	\usepackage{bm}
	\usepackage[utf8]{inputenc}
	\usepackage[T1]{fontenc}
	\usepackage{mathptmx}
	\usepackage{microtype}
	\microtypesetup{expansion=false}

	\usepackage{tikz,pgf,placeins,subcaption}
	\usetikzlibrary{calc}
	\graphicspath{{img/}}

	\usepackage{amsfonts}
	\usepackage{amsmath}
	\usepackage{amssymb}
	\usepackage{dcolumn}
	\usepackage{enumerate}
	\usepackage[version=4]{mhchem}
	\usepackage{siunitx}
		\DeclareSIUnit\atom{at}
		\DeclareSIUnit\sccm{sccm}
		\DeclareSIUnit\torr{torr}
		\DeclareSIUnit\deg{\ensuremath{^\circ}}
	\usepackage{soul}
	\usepackage{xcolor}
	
	\newcommand*{\B}[1]{\relax\ifmmode\bm{#1}\else\textbf{#1}\fi}

	\usepackage{graphicx}

	\usepackage[colorlinks,allcolors=blue]{hyperref}
	\usepackage[capitalize]{cleveref} % Must be loaded after hyperref
	
	\definecolor{Blue}{RGB}{0,0,220}

	\definecolor{Green}{RGB}{0,120,0}

	\definecolor{Purple}{RGB}{220,0,220}
	
	\definecolor{Red}{RGB}{220,0,0}

	\makeatletter
	\let\@fnsymbol\@fnsymbol@latex
	\@booleanfalse\altaffilletter@sw
	\makeatother

\begin{document}
\title[Superconducting \ce{V3Si} for quantum circuit applications]{Superconducting \texorpdfstring{\ce{V3Si}}{V3Si} for quantum circuit applications}

\author{T.D. Vethaak}
\affiliation{Université Grenoble Alpes, CEA, Grenoble INP, IRIG, PHELIQS, 38000 Grenoble, France}
\affiliation{Université Grenoble Alpes, CEA LETI, 38000 Grenoble, France}
\author{F. Gustavo}
\affiliation{Université Grenoble Alpes, CEA, Grenoble INP, IRIG, PHELIQS, 38000 Grenoble, France}
\author{T. Farjot}
\affiliation{Université Grenoble Alpes, CEA LETI, 38000 Grenoble, France}
\author{T. Kubart}
\affiliation{Division of Solid-State Electronics, Department of Electrical Engineering, Uppsala University, 75103 Uppsala, Sweden.}
\author{P. Gergaud}
\affiliation{Université Grenoble Alpes, CEA LETI, 38000 Grenoble, France}
\author{S-L. Zhang}
\affiliation{Division of Solid-State Electronics, Department of Electrical Engineering, Uppsala University, 75103 Uppsala, Sweden.}
\author{F. Lefloch}
\affiliation{Université Grenoble Alpes, CEA, Grenoble INP, IRIG, PHELIQS, 38000 Grenoble, France}
\author{F. Nemouchi}
\email[ Corresponding author: ]{Fabrice.Nemouchi@cea.fr}
\affiliation{Université Grenoble Alpes, CEA LETI, 38000 Grenoble, France}

\date\today

\begin{abstract}
	\noindent \ce{V3Si} thin films are known to be superconducting with transition temperatures up to \SI{15}{\kelvin}, depending on the annealing temperature and the properties of the substrate underneath.
	Here we investigate the film structural properties with the prospect of further integration in silicon technology for quantum circuits.    
	Two challenges have been identified: (i) the large difference in thermal expansion rate between \ce{V3Si} and the Si substrate leads to large thermal strains after thermal processing, and (ii) the undesired silicide phase \ce{VSi2} forms when \ce{V3Si} is deposited on silicon.
	The first of these is studied by depositing layers of \SI{200}{\nano\meter} \ce{V3Si} on wafers of sapphire and oxidized silicon, neither of which react with the silicide.
	These samples are then heated and cooled between room temperature and \SI{860}{\celsius}, during which in-situ XRD measurements are performed.
	Analysis reveals a highly non-linear stress development during heating with contributions from crystallization and subsequent grain growth, as well as the thermal expansion mismatch between silicide and substrate, while the film behaves thermoelastically during cooling.
	The second challenge is explored by depositing films of 20, 50, 100 and \SI{200}{\nano\meter} of \ce{V3Si} on bulk silicon.
	For each thickness, six samples are prepared, which are then annealed at temperatures between 500 and \SI{750}{\celsius}, followed by measurements of their resistivity, residual resistance ratio and superconducting critical temperature.
	A process window is identified for silicide thicknesses of at least \SI{100}{\nano\meter}, within which a trade-off needs to be made between the quality of the \ce{V3Si} film and its consumption by the formation of \ce{VSi2}.
\end{abstract}

\maketitle

\section{Introduction}
	
	Quantum technologies based on solid state matter are now spreading from academic laboratories to the most advanced semiconductor foundries~\cite{arute2019quantum,zwerver2021qubits}.
	Among the materials to be integrated, superconducting thin films are of major interest as they are part of many quantum architectures.
	For instance, such films can be used to manipulate, read and couple superconducting or spin qubits with the use of superconducting resonators~\cite{majer2007coupling,mi2018coherent}, are developed for non-dissipative interconnections or can even be part of the quantum device itself when integrated as source and drain contacts of CMOS transistors.
	Since silicon technology is by far the most advanced method of nanofabrication, it is important to identify superconducting materials that are fully compatible with such an environment.
	Among these, silicides appears as the most suitable materials~\cite{shibata1981optimally,zhang2014metal}, and as \ce{V3Si} has the highest known critical temperature of any silicide known so far~\cite{hardy1953superconducting,blumberg1960correlations}, it is a prime candidate for these applications.
	In working towards integration of \ce{V3Si}, two key problems have been identified: (i) a reduction in the superconducting critical temperature due to increased thermal strain after annealing~\cite{vethaak2021influence}, and (ii) the formation of the non-superconducting phase \ce{VSi2} at the interface with a silicon channel or substrate.
	
	In systems where reservoirs of both Si and V are available, the undesired phase \ce{VSi2} is expected to form, which is both the first phase to nucleate~\cite{krautle1974kinetics} and is thermodynamically favored~\cite{pretorius1993thin}.
	This rules out the self-aligned silicide (SALICIDE) process for the formation of the desired \ce{V3Si}, in which pure metal is deposited on exposed silicon contacts.
	Instead, it has motivated the adoption of \ce{V3Si} sputtering from compound targets~\cite{michikami1982v3si,theuerer1964getter}, after which an annealing step is required to trigger the crystallization.
	This thermal processing leads to the build-up of strain in the final silicide film, which is known to negatively affect its critical temperature~\cite{testardi1970unusual,testardi1971unusual95,testardi1971unusual,testardi1972structural,batterman1964crystal,batterman1966low}.
	Earlier work has shown that this strain depends strongly on the mismatch in thermal expansion coefficient (TEC) between the silicide and the substrate, and can lead to a suppression in superconducting critical temperature $T_\text{c}$ by \SI{2}{\kelvin} on silicon~\cite{vethaak2021influence}.
	Section~\ref{sec:xrd} provides a detailed analysis of the impact of crystallization, grain growth and thermal expansion mismatch on the strongly temperature-dependent development of stress in \ce{V3Si} thin films.
	
	Besides causing the buildup of large tensile stresses, thermal processing of \ce{V3Si} on silicon substrates also leads to the formation of \ce{VSi2}.
	The presence of \ce{VSi2} at the interface between the superconducting \ce{V3Si} and the Si substrate may not be a problem per se.
	It is imperative, however, that a layer of \ce{V3Si} with a thickness on the order of its superconducting coherence length remains.
	The simultaneous formation of the two silicide phases after the deposition of amorphous \ce{V3Si} on bulk silicon wafers is discussed in section~\ref{sec:v3si_vsi2}.
	We find that when \SI{50}{\nano\meter} or less of \ce{V3Si} is deposited, no superconductivity is observed after crystallization annealing.
	For layers of 100 or \SI{200}{\nano\meter}, the superconducting critical temperature of the film is found to initially improve with annealing temperature, until at some point superconductivity disappears when all \ce{V3Si} is consumed by a growing layer of \ce{VSi2}.
	
\section{\label{sec:xrd}In-situ XRD analysis}\FloatBarrier
	
	Sputtering a \ce{V3Si} target leads to the deposition of an amorphous layer of the desired intermetallic compound which presents superconductivity at temperatures of around \SI{1}{\kelvin}, far below that of the crystalline form (measured but not shown).
	In order to nucleate the crystalline \ce{V3Si} phase, heat has to be applied.
	During the experiment described in the present section, the \ce{V3Si} was isolated from any potential element (V and Si especially) that would affect the crystallization phenomenon.
	Thus, depositing the intermetallic compound on oxides (silica and sapphire) offers the possibility to study the isolated film by itself, affected by the substrate only mechanically.	
	The evolution of the stresses in the \ce{V3Si} film as a function of the annealing temperature was monitored by in-situ X-ray diffraction (XRD) using a Smartlab Rigaku X-ray diffractometer equipped with a copper (Cu \ce{K_{\alpha}}) source, a punctual detector and a DHS 1100 Anton Paar furnace using a nitrogen atmosphere at a pressure slightly above ambient, i.e. \SI{1.25}{\bar}, to avoid atmospheric contamination.
	
\begin{figure}
		\centering
		\includegraphics[width=\columnwidth]{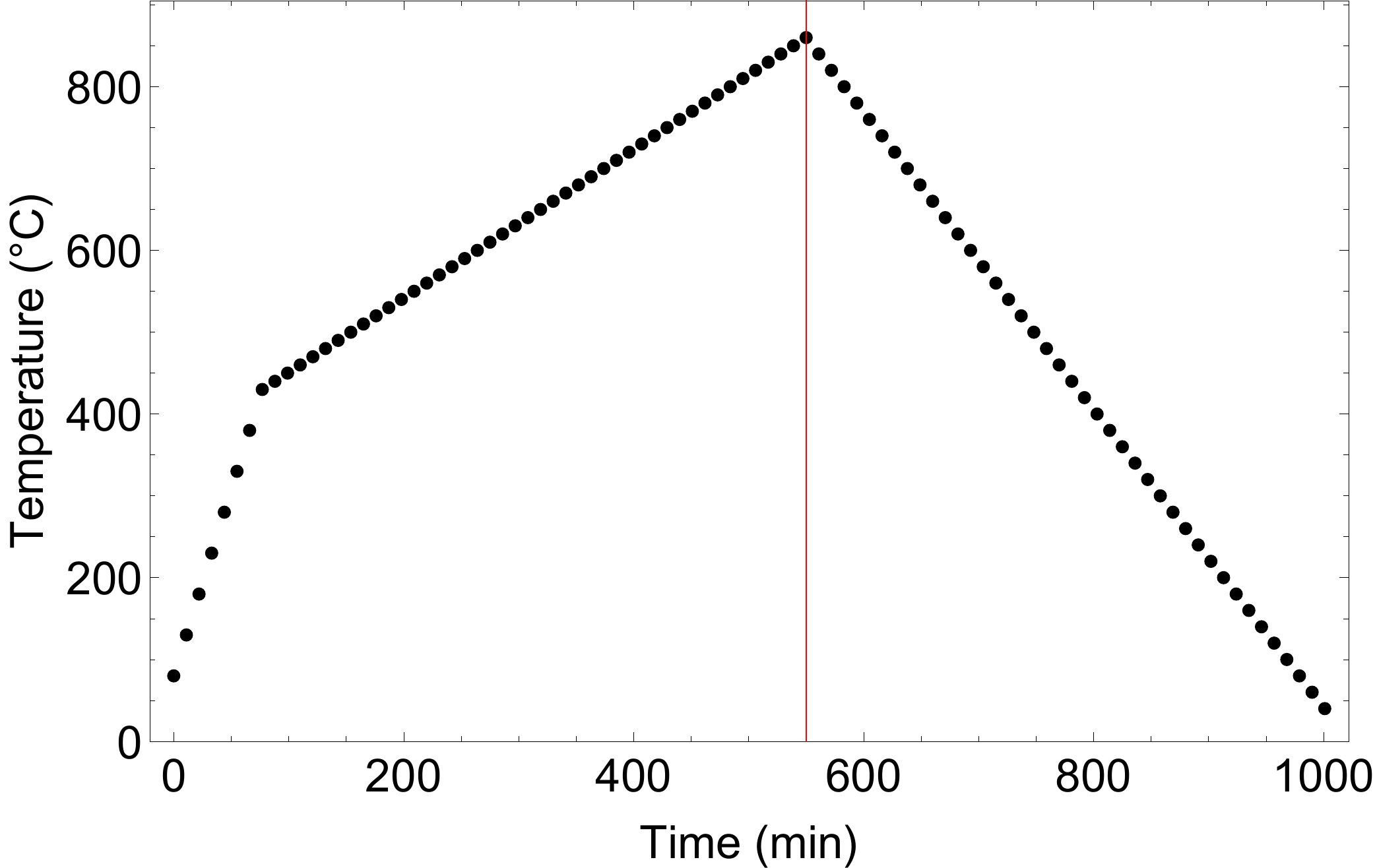}
		\caption{\label{fig:figiii1}Temperature profile versus annealing time applied during the in-situ XRD experiments. The red line in this and the following graphs in this section indicates the maximum temperature reached, after which cooling starts.}
	\end{figure}

	Layers of \SI{200}{\nano\meter} \ce{V3Si} were deposited on two types of substrates: oxidized silicon with a \SI{20}{\nano\meter} thick surface oxide, and sapphire.
	Both in-plane and out-of-plane XRD measurements were performed on different samples of each substrate.
	For both measurements, a parallel slit analyzer with an angular aperture of \SI{0.114}{\deg} for out-of-plane measurement and \SI{1}{\deg} for in-plane ones, is positioned in front of the detector in order to be insensitive to peak shift due to the radial displacement of the sample induced by thermal expansion of the sample holder.
	This way, the observed peak shift is only related to a real d-spacing variation.
	For both measurements, $2\theta$--$\omega$ scans around the \ce{V3Si} (210) close to $2\theta=\SI{43}{\deg}$ were performed.
	Profiles were acquired while heating the samples from \SI{50}{\celsius} up to \SI{860}{\celsius} and then during cooling from \SI{860}{\celsius} to room temperature.
	The temperature profile versus time is given in Fig.~\ref{fig:figiii1}.
	
\begin{figure}
		\centering
		\includegraphics[width=\columnwidth]{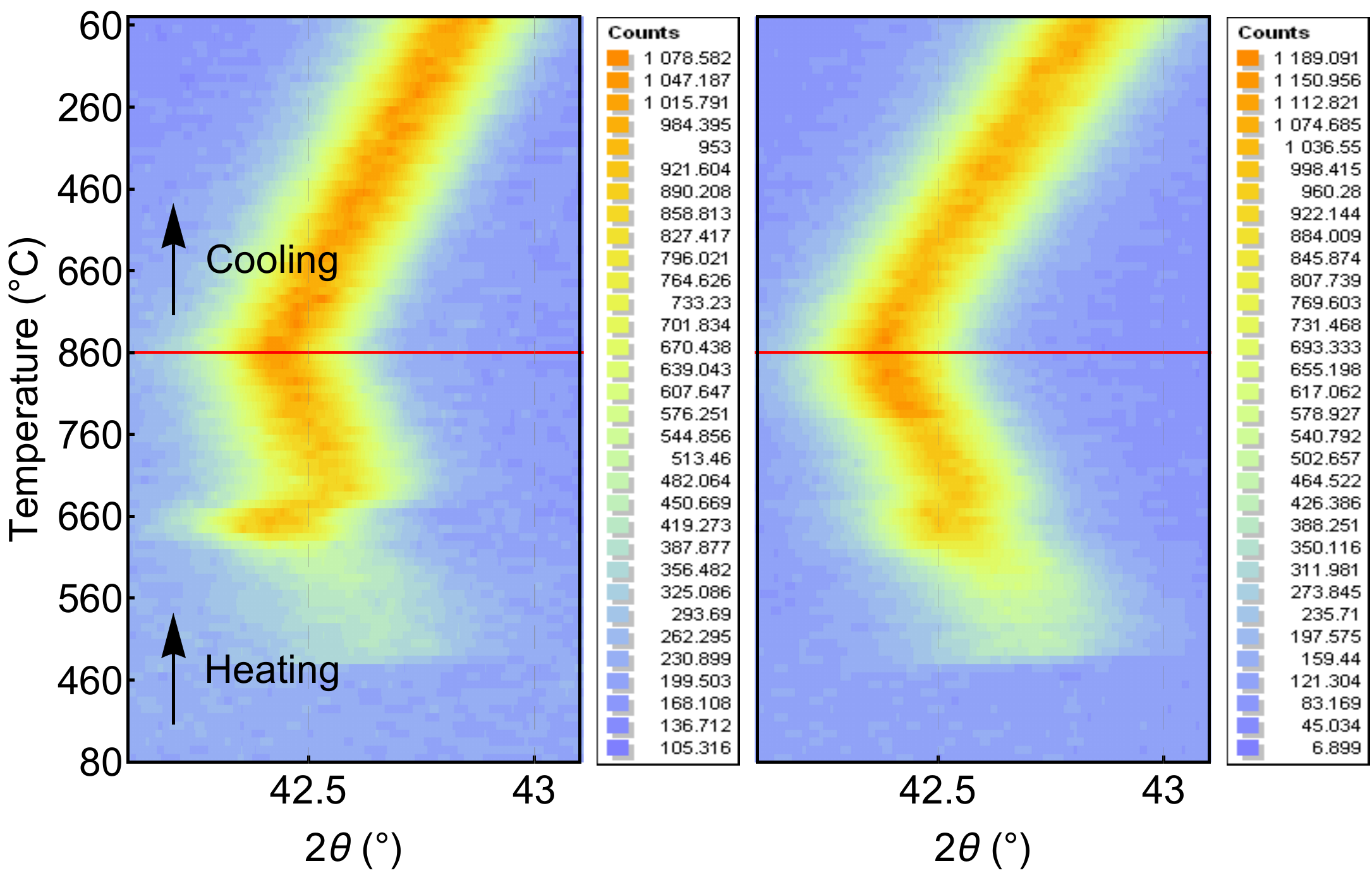}
		\caption{\label{fig:figiii2}In-situ out-of-plane XRD patterns measured during the \ce{V3Si} crystallization on \B{(Left)} a sapphire substrate and \B{(Right)} a silicon substrate.}
	\end{figure}

\begin{figure}
		\centering
		\includegraphics[width=\columnwidth]{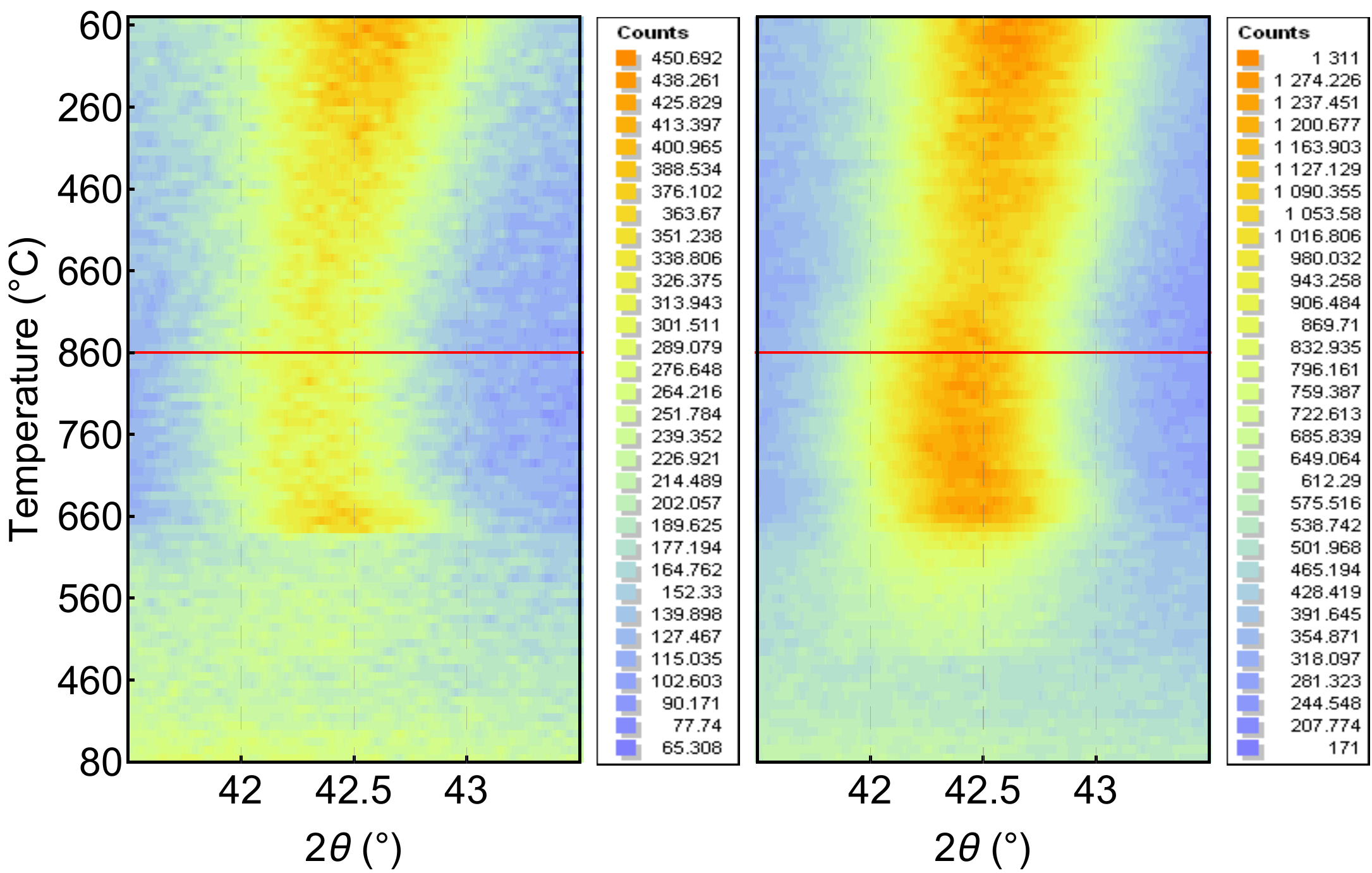}
		\caption{\label{fig:figiii3}In-situ in-plane XRD patterns measured during the \ce{V3Si} crystallization on \B{(Left)} a sapphire substrate and \B{(Right)} a silicon substrate.}
	\end{figure}

	Peak fitting was performed with the HighScore Plus software from PANalytical, from which the d-spacing and the integral breadth of the \ce{V3Si} (210) peak were considered.
	The \ce{V3Si} crystallization followed by in-situ XRD is shown as a contour map for the out-of-plane measurements in Fig.~\ref{fig:figiii2} and the in-plane measurements in Fig.~\ref{fig:figiii3}.
	
\begin{figure}
		\centering
		\includegraphics[width=\columnwidth]{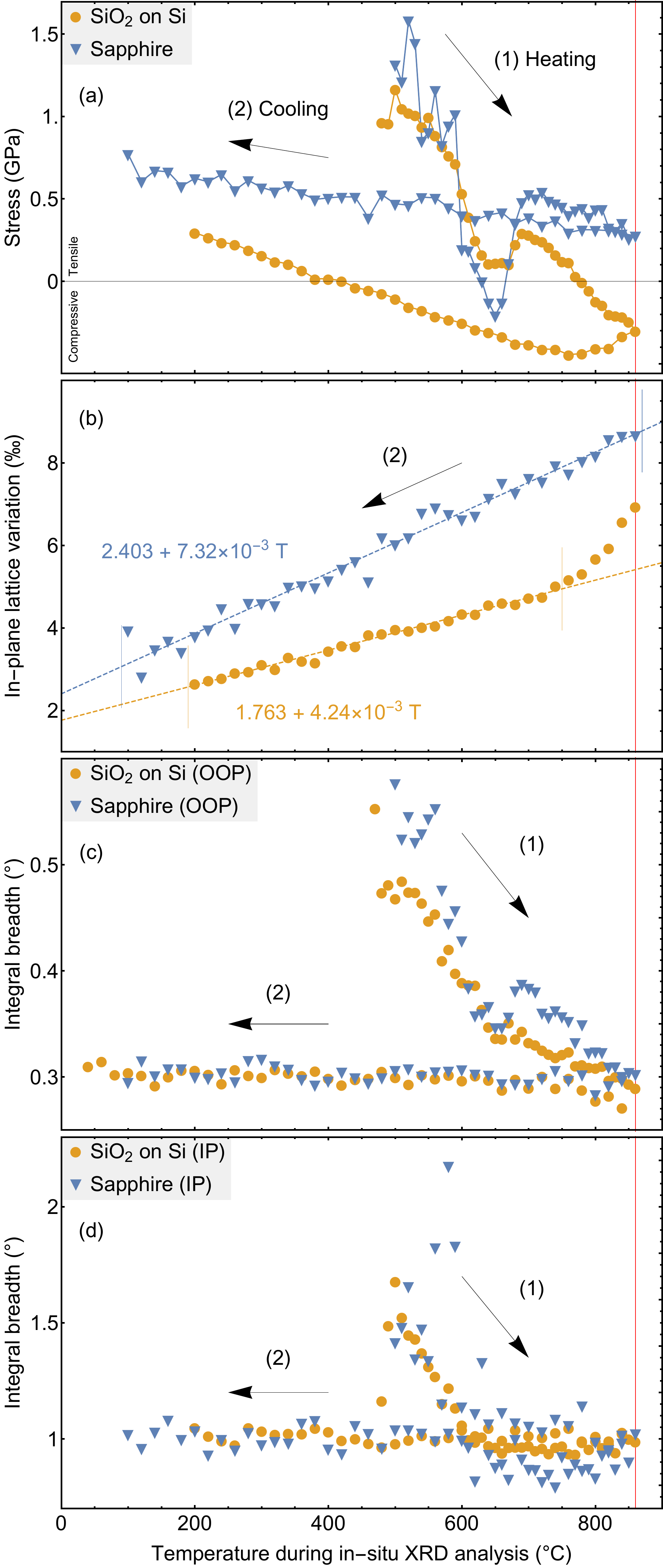}
		\caption{\label{fig:strain}\B{(a)} Stress evolution versus annealing temperature for both the sapphire and the silicon substrates. \B{(b)} Strain of the \ce{V3Si} layer versus the annealing temperature during cooling. \B{(c,d)} Out-of-plane and in-plane integral breadth of the \ce{V3Si} (210) peak versus annealing temperature.}
	\end{figure}

	During heating, for both substrates, a characteristic peak of (210) \ce{V3Si} is observed above \SI{500}{\celsius}, which marks the crystallization temperature of the as-deposited amorphous \ce{V3Si} layer.
	One can thus deduce that the nucleation temperature of \ce{V3Si} crystalline phase is close to 500 °C.
	From in-plane and out-of-plane peak positions, one can extract the stress and the temperature-dependent stress-free lattice parameter of the layer.
	Assuming a bi-axial state of stress in the layer, it can be deduced following the $\sin^2\psi$ methodology~\cite{noyan2013residual},
	\begin{equation}\epsilon_\psi=\dfrac{(d_\psi-d_0)}{d_0}=\dfrac{1}{2}S_2(hkl)\sigma\sin^2\psi+2S_1(hkl)\sigma,\end{equation}
	where $\sigma$ is the in-plane bi-axial state of stress, $\epsilon_\psi$ and $d_\psi$ are the strain and the $d$-spacing along the direction $\psi$, respectively, $d_0$ is the stress-free $d$-spacing and $\psi$ is the angle between the considered direction and the normal to the surface. 
	$\tfrac{1}{2}S_2(hkl)$ and $S_1(hkl)$ are the X-ray Elastic Constants (XEC) for the hkl plane considered.
	The XEC can be easily calculated~\cite{murray2013equivalence} by knowing the single-crystal elastic constants of the \ce{V3Si} structure~\cite{gaillac2016elate}.
	For the (210) plane, using the arithmetic average of the Voigt and Reuss XEC, known as the Neerfeld limit, we found $\tfrac{1}{2} S_2(210) = \SI{6.69E-6}{\per\mega\pascal}$ and $S_1(210) = \SI{-1.66E-6}{\per\mega\pascal}$.
	The in-plane stress is then given by~\cite{noyan2013residual}
	\begin{equation}\sigma=\dfrac{2}{S_2(hkl)}\left(\dfrac{d_\parallel-d_\perp}{d_0}\right),\end{equation}
	and the stress-free $d$-spacing, $d_0$, by
	\begin{equation}d_0=\dfrac{d_\perp-Ad_\parallel}{1-A},\quad\text{where}\quad A=\dfrac{4S_1(hkl)}{S_2(hkl)+4S_1(hkl)}.\end{equation}
	In Fig.~\ref{fig:strain}a, the stress evolution deduced from this analysis is plotted versus the annealing temperature for both types of substrate.
	In both cases, the initial stress at crystallization is highly tensile, around $+\SI{1500}{\mega\pascal}$.
	This is in agreement with the expected volume shrinkage that occurs during the crystallization, assuming that the density of the amorphous phase is smaller than the density of the crystalline phase.
	This tensile stress is then partially relaxed.

	In Fig.~\ref{fig:strain}c,d the integral breadth (IB) evolution of the (210) peak vs annealing temperature is plotted, which can be correlated to the crystallite size through the Scherrer equation~\cite{scherrer1918nachr,cullity2001elements}.
	This figure shows that the IB is high at the crystallization temperature and then goes down with increasing temperature, a behavior that is similar for the in-plane and out-of-plane directions, and can be interpreted as a crystallite size increase.
	However, one should note that the in-plane and and out-of-plane setup was different and can not be compared directly.
	
	Indeed, for the in-plane setup, the IB decreased down to 1° at 650°C and stayed constant during further annealing.
	This value of 1° is the instrumental width given by a silicon reference powder, which implies that the in-plane crystallite size is very large above \SI{650}{\celsius} and can no longer be estimated from IB broadening.
	As we observe an increase of the crystallite size out of plane up to \SI{860}{\celsius}, we may assume that the lateral crystallite size also continues to grow up to this temperature.
	In the case of the out-of-plane measurements, the setup was more accurate, and the estimated crystallite size is about \SI{30}{\nano\meter} for the highest annealing temperature.
	It can clearly be concluded that there is a strong anisotropy in grain size on both substrates.
	On both substrates, the in-plane IB stabilizes at \SI{650}{\celsius} due to saturation of the instrument resolution, while the detected out-of-plane crystallite size increases continuously to the maximum temperature of the experiment.
	
	During cooling, the stress evolution exhibits a linear behavior, in agreement with a pure thermoelastic stress induced by the TEC mismatch between the film and the substrate.
	This is confirmed by looking at the in-plane lattice parameter evolution with temperature (Fig~\ref{fig:strain}b).
	The behavior is linear as expected, but more interestingly, the slope of these straight lines is \SI{7.3E-6}{\per\celsius} and \SI{4.2E-6}{\per\celsius} for the sapphire and Si substrate respectively, in good agreement with the in-plane TEC of (0001) sapphire ($\SI{6.5E-6}{}<\alpha<\SI{7.6E-6}{\per\celsius}$ in the temperature range 300--\SI{800}{\celsius})~\cite{vodenitcharova2006effect}, and with the TEC of Si ($\SI{3.6E-6}{}<\alpha<\SI{4.3E-6}{\per\celsius}$ in the temperature range 200--\SI{800}{\celsius})~\cite{watanabe2004linear}.
	From the stress-free lattice parameter extraction, we calculated a TEC of the \ce{V3Si} phase of \SI{9.2E-6}{\per\celsius}.
	The TEC mismatch between the film and the substrate should therefore have added a tensile stress to the stress measured at \SI{860}{\celsius}, in agreement with the observation in Fig.~\ref{fig:strain}a.
	As expected, since the TEC mismatch is larger with the Si substrate than with the sapphire, the tensile stress increases faster on Si than on sapphire.
	Extrapolating this thermoelastic behavior to the device operating temperature (i.e. a few K), a stress level of \SI{0.87\pm0.07}{\giga\pascal} is obtained for \ce{V3Si} on sapphire and \SI{0.91\pm0.03}{\giga\pascal} for \ce{V3Si} on silicon.
	Of course these values depend on the initial residual stress before cooling, and thus on the thermal budget applied to the samples.
	Since the slow cooling occurred entirely thermoelastically and followed the TEC of the substrate, it is not expected that the rate of cooling has any impact on the final stress obtained.
	The difference in room-temperature strain between this report and those found in earlier experiments where in-situ XRD analysis was performed up to \SI{1000}{\celsius}~\cite{vethaak2021influence} is thus attributed to differences in the maximum temperature that was reached.

\section{\label{sec:v3si_vsi2}Competition between \texorpdfstring{\ce{V3Si}}{V3Si} and \texorpdfstring{\ce{VSi2}}{VSi2} formation}

	A second set of experiments was performed to study the stability of the superconducting \ce{V3Si} phase on a silicon substrate.
	After sputter deposition of \ce{V3Si} on a silicon substrate, an intermixing layer of Si and V can be expected to form where the atomic concentration of vanadium ranges from zero in the substrate to $3/4$ within the deposited layer.
	Above this mixed zone, the precise matching of the deposited atomic ratio to the stoichiometry of \ce{V3Si} will prevent the formation of \ce{VSi2}.
	Within the mixed zone however, \ce{VSi2} nucleation is likely to occur.
	Since this mixed zone is interfaced from below by pure silicon, and from above by \ce{V3Si}, its disappearance or consumption by \ce{V3Si} formation would require that V move upwards while leaving progressively purified Si behind.
	Such movement is prevented by the chemical potential, as the thermodynamically stable state in the presence of excess silicon is for the vanadium to be bound in \ce{VSi2}~\cite{pretorius1993thin}.
	Furthermore, even if a V reservoir were present in the form of a super-stoichiometric vanadium concentration (i.e. above 75\%) in the deposited layer, it is unlikely that the mixed zone would be consumed by \ce{V3Si} formation.
	Since silicon is the dominant diffusing species in \ce{VSi2} and at the V/Si interface, with diffusion rates of two orders of magnitude higher than that of V in \ce{V3Si}~\cite{chu1974identification,schutz1979formation}, any part of the intermixing layer with lower vanadium concentration than that of \ce{V3Si} is expected to quickly extend downward and decrease in V content due to upward Si diffusion, preventing \ce{V3Si} growth and aiding the formation of \ce{VSi2}.
	
\begin{figure}
		\centering
		\begin{tikzpicture}
			\node[anchor=north west,inner sep=0] at (0,0) {
				\includegraphics[width=\columnwidth]{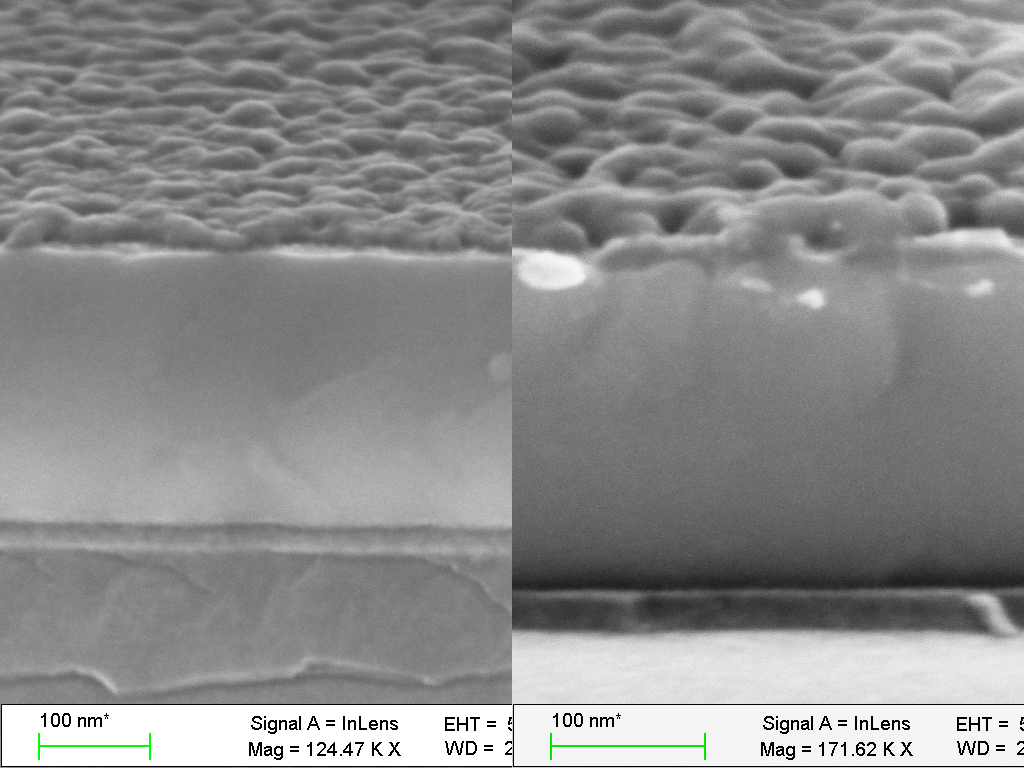}
				};
			\node[color=black,anchor=north west,yshift=-0.1cm] at (0.12,-0.01) {\Large \B{600°C}};
			\node[color=white,anchor=north west,yshift=-0.1cm] at (0.1,0) {\Large \B{600°C}};
			\node[color=black,anchor=north west,yshift=-0.1cm] at ($(0.5\columnwidth,0)+(0.12,-0.01)$) {\Large \B{650°C}};
			\node[color=white,anchor=north west,yshift=-0.1cm] at ($(0.5\columnwidth,0)+(0.1,0)$) {\Large \B{650°C}};
			
			\draw[thick, red] (0.3,-4.4) -- (1.3,-4.43);
			\draw[thick, red] (0.3,-4.62) -- (1.3,-4.65);
			\node[black] at (2.0,-3.3) {\large \B{\ce{V3Si}}};
			\node[white] at (2.02,-3.31) {\large \B{\ce{V3Si}}};
			\node[black] at (2.0,-4.54) {\large \B{\ce{SiO2}}};
			\node[white] at (2.02,-4.55) {\large \B{\ce{SiO2}}};
			\node[black] at (2.0,-5.4) {\large \B{\ce{Si}}};
			\node[white] at (2.02,-5.41) {\large \B{\ce{Si}}};
			
			\draw[thick, red] ($(0.5\columnwidth,0)+(0.3,-5.0)$) -- ($(0.5\columnwidth,0)+(1.3,-5.0)$);
			\draw[thick, red] ($(0.5\columnwidth,0)+(0.3,-5.33)$) -- ($(0.5\columnwidth,0)+(1.3,-5.33)$);
			\node[black] at ($(0.5\columnwidth,0)+(2.0,-3.66)$) {\large \B{\ce{V3Si}}};
			\node[white] at ($(0.5\columnwidth,0)+(2.02,-3.67)$) {\large \B{\ce{V3Si}}};
			\node[black] at ($(0.5\columnwidth,0)+(2.0,-5.16)$) {\large \B{\ce{SiO2}}};
			\node[white] at ($(0.5\columnwidth,0)+(2.02,-5.17)$) {\large \B{\ce{SiO2}}};
			\node[white] at ($(0.5\columnwidth,0)+(2.0,-5.66)$) {\large \B{\ce{Si}}};
			\node[black] at ($(0.5\columnwidth,0)+(2.02,-5.67)$) {\large \B{\ce{Si}}};
		\end{tikzpicture}
		\begin{tikzpicture}
			\node[anchor=north west,inner sep=0] at (0,0) {
				\includegraphics[width=\columnwidth]{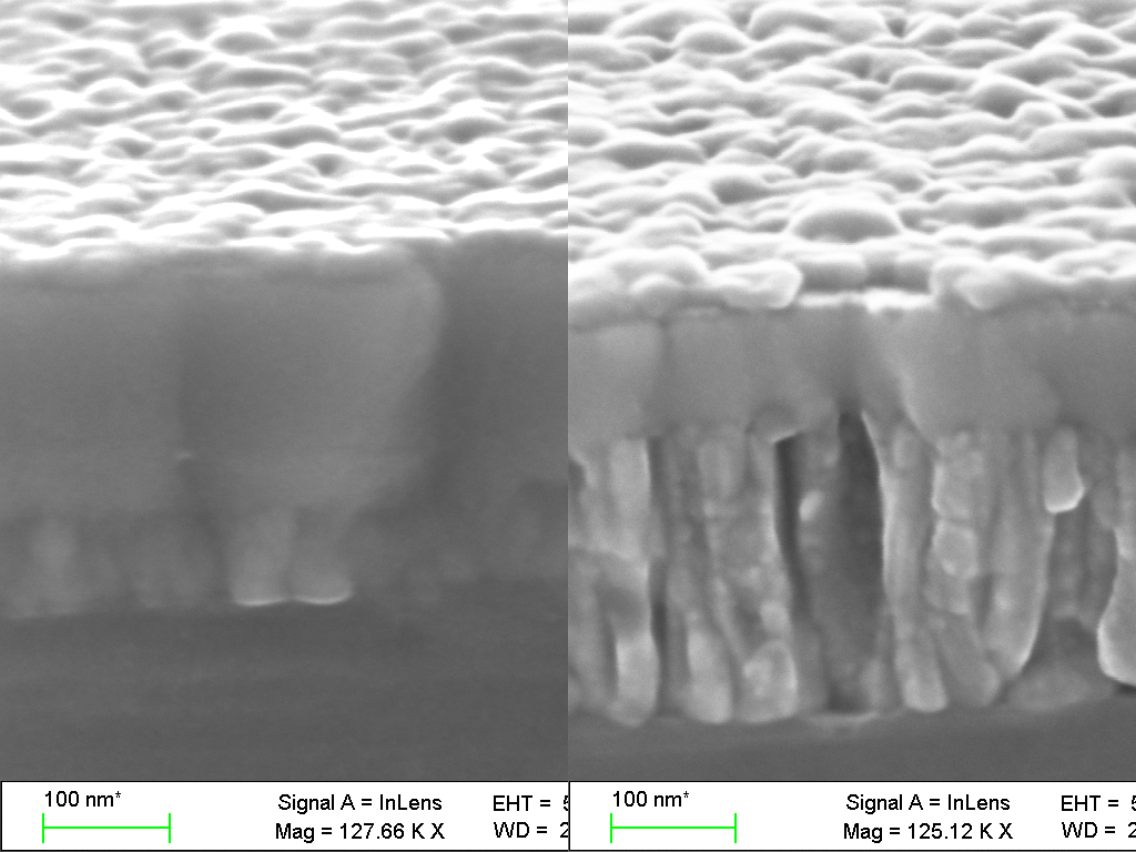}
			};
			\node[color=white,anchor=north west,yshift=-0.1cm] at (0.12,-0.01) {\Large \B{600°C}};
			\node[color=black,anchor=north west,yshift=-0.1cm] at (0.1,0) {\Large \B{600°C}};
			\node[color=white,anchor=north west,yshift=-0.1cm] at ($(0.5\columnwidth,0)+(0.12,-0.01)$) {\Large \B{650°C}};
			\node[color=black,anchor=north west,yshift=-0.1cm] at ($(0.5\columnwidth,0)+(0.1,0)$) {\Large \B{650°C}};
			
			\draw[thick, red] (0.3,-3.9) -- (1.3,-3.9);
			\draw[thick, red] (0.3,-4.66) -- (1.3,-4.66);
			\node[black] at (2.0,-3.1) {\large \B{\ce{V3Si}}};
			\node[white] at (2.02,-3.11) {\large \B{\ce{V3Si}}};
			\node[black] at (2.0,-4.34) {\large \B{\ce{VSi2}}};
			\node[white] at (2.02,-4.35) {\large \B{\ce{VSi2}}};
			\node[black] at (2.0,-5.3) {\large \B{\ce{Si}}};
			\node[white] at (2.02,-5.31) {\large \B{\ce{Si}}};
			
			\draw[thick, red] ($(0.5\columnwidth,0)+(0.3,-3.3)$) -- ($(0.5\columnwidth,0)+(1.3,-3.3)$);
			\draw[thick, red] ($(0.5\columnwidth,0)+(0.3,-5.5)$) -- ($(0.5\columnwidth,0)+(1.3,-5.5)$);
			\node[black] at ($(0.5\columnwidth,0)+(2.0,-2.86)$) {\large \B{\ce{V3Si}}};
			\node[white] at ($(0.5\columnwidth,0)+(2.02,-2.87)$) {\large \B{\ce{V3Si}}};
			\node[black] at ($(0.5\columnwidth,0)+(2.0,-4.46)$) {\large \B{\ce{VSi2}}};
			\node[white] at ($(0.5\columnwidth,0)+(2.02,-4.47)$) {\large \B{\ce{VSi2}}};
			\node[black] at ($(0.5\columnwidth,0)+(2.0,-5.76)$) {\large \B{\ce{Si}}};
			\node[white] at ($(0.5\columnwidth,0)+(2.02,-5.77)$) {\large \B{\ce{Si}}};
		\end{tikzpicture}
		\caption{\label{fig:sem_sio2_si}\B{(Top)} A \SI{200}{\nano\meter} layer of \ce{V3Si} is deposited on a silicon substrate with \SI{20}{\nano\meter} of thermal oxide and is annealed at \SI{600}{\celsius} (left) and \SI{650}{\celsius} (right). No formation of \ce{VSi2} is observed. \B{(Bottom)} When an equal thickness of \ce{V3Si} is deposited on HF-cleaned silicon and annealed at \SI{600}{\celsius} (left) and \SI{650}{\celsius} (right), a \ce{VSi2} layer with distinct morphology appears.}
	\end{figure}
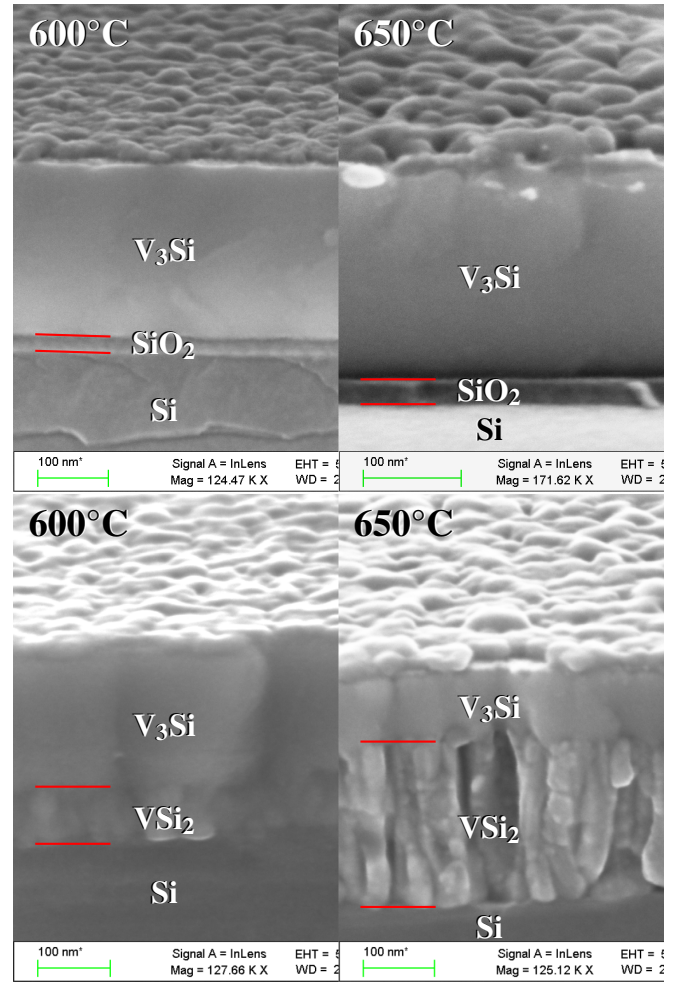

	Once \ce{VSi2} and \ce{V3Si} are both present, competition will arise between the further growth of either silicide.
	While \ce{V3Si} (\ce{V_{0.75}Si_{0.25}}) has a larger effective heat of formation (EHF) from pure V and Si per mole of atoms involved in total~\cite{pretorius1993thin}, the energy gain \emph{per vanadium atom} is greater for \ce{VSi2} (\ce{V_{0.33}Si_{0.67}}), of which three times as many molecules can be formed for a fixed amount of vanadium.
	When \ce{V3Si} interfaces with silicon, it is therefore energetically favorable for the following reaction to occur (EHF indicated below),
	\begin{equation}\label{eq:v3si_vsi2_reaction}\underbrace{\ce{V3Si}}_{\SI{-45.2}{\kilo\joule/\mole}\text{ at}} + \underbrace{\phantom{\!\!\!\!_0}5\ce{Si}\phantom{\!\!\!\!_0}}_{\phantom{\si{\per\mole}}0\phantom{\si{\per\mole}}} \rightarrow \underbrace{3\ce{VSi2}}_{\SI{-40.2}{\kilo\joule/\mole}\text{ at}},\end{equation}
	where an energy of $\SI{181}{\kilo\joule/\mole}$ is gained per mole of \ce{V3Si} that is transformed.
	Scanning electron microscope (SEM) images of the growth of the \ce{VSi2} phase during thermal processing are shown in Fig.~\ref{fig:sem_sio2_si}.

\begin{figure}
		\centering
		\includegraphics[width=\columnwidth]{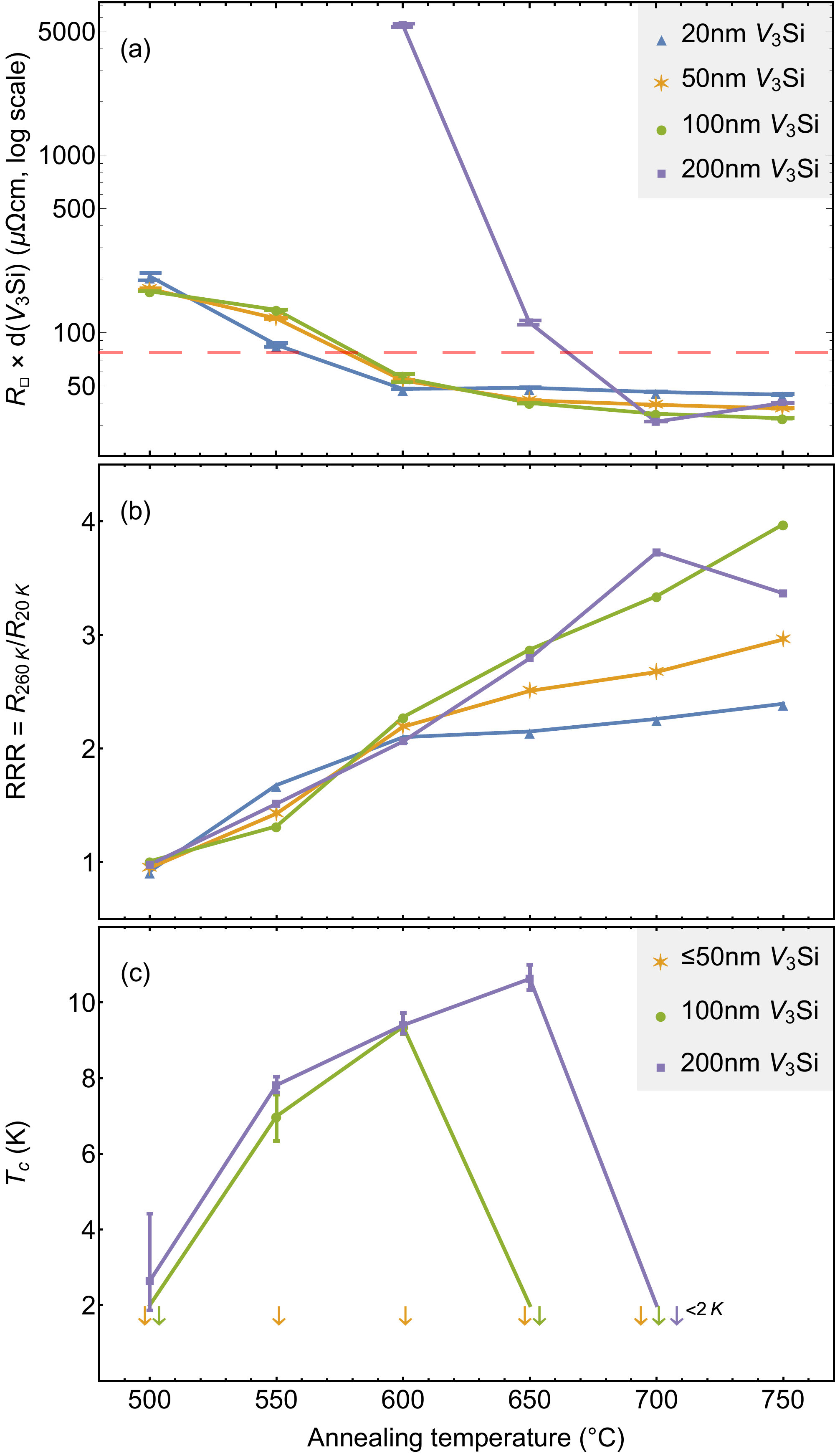}
		\caption{\label{fig:v3sicombinedplot}\B{(a)} The resistivity of each annealed sample, the red dashed line indicates the lowest resistivity expected for \ce{V3Si} (see main text). \B{(b)} The residual resistance ratio (RRR). \B{(c)} The critical temperature, taken to be the point where 50\% of the normal-state resistance is lost, with error bars indicating temperatures at which 10\% and 90\% of normal-state resistance is observed. Arrows down ($\downarrow$) indicate that the critical temperature (if any) is below \SI{2}{\kelvin}.}
	\end{figure}

	To study this interplay, a different set of samples was prepared in addition to those discussed in section~\ref{sec:xrd}.
	Layers with thicknesses of 20, 50, 100 and \SI{200}{\nano\meter} of \ce{V3Si} were deposited onto \SI{200}{\milli\meter} (100)-oriented silicon wafers from a compound \ce{V3Si} target using RF magnetron sputtering equipment.
	This target had a measured silicon content of between $22.7\pm0.2\text{ at.}\%$ (WDXRF) and $25.7\pm0.2\text{ at.}\%$ (supplier data), which could further differ from the final Si content in the deposited layer by $1\text{ at.}\%$~\cite{theuerer1964getter}.
	These wafers were cleaned with hydrofluoric acid (HF) before entering the sputtering chamber.
	Smaller pieces of $2\times\SI{2}{\centi\meter\squared}$ were then annealed during 2 minutes in a Jipelec furnace with a \SI{20}{\celsius/\minute} ramp rate.
	For each thickness of \ce{V3Si}, six different samples were thermally processed at temperatures between 500 and \SI{750}{\celsius}, as indicated in Fig.~\ref{fig:v3sicombinedplot}.
	These samples were then characterized by measuring their room-temperature sheet resistance (corrected for their shape and size by calculating the sheet resistance ratio of multiple unannealed wafer-sized and $2\times\SI{2}{\centi\meter\squared}$ samples), and by following the resistance of smaller samples during slow cooling to cryogenic temperatures of \SI{2}{\kelvin} in order to extract both the superconducting critical temperature and the residual resistance ratio (RRR).
	
	At \SI{500}{\celsius}, the resistivity of the thinner layers (20, 50 and \SI{100}{\nano\meter}) was found to be comparable to that of as-deposited amorphous \ce{V3Si} (\SI{180\pm10}{\micro\ohm\centi\meter}), while it exceeded this value by multiple orders of magnitude for the thicker layer of \SI{200}{\nano\meter} (out of range for our instruments), as shown in Fig.~\ref{fig:v3sicombinedplot}a.
	For all thicknesses, the resistivity then decreased with annealing temperature, to values below that measured for a reference sample with \SI{200}{\nano\meter} \ce{V3Si} on non-reacting sapphire annealed at \SI{900}{\celsius} (\SI{77.1}{\micro\ohm\centi\meter}, indicated as a red dashed line).
	Since this reference value is the lowest resistivity that we have found for any layer of crystallized \ce{V3Si}, in a sample where we confirmed by XRD measurements that no other vanadium-rich compound was formed, we take any value lower than this on silicon substrates to indicate the presence of \ce{VSi2}.
	The formation of \ce{VSi2} was confirmed by XRD analysis and identified at the Si/\ce{V3Si} interface by SEM imaging (see Fig.~\ref{fig:sem_sio2_si}, bottom).

	Smaller samples of $4\times\SI{10}{\milli\meter}$ were then wire bonded for four-point resistance measurements during both cooling and heating between 300 and \SI{2}{\kelvin} in a Physical Properties Measurement System (PPMS).
	The residual resistance ratio of each sample was calculated as the ratio between the resistances at \SI{260}{\kelvin} and \SI{20}{\kelvin}, shown in Fig.~\ref{fig:v3sicombinedplot}b.
	This ratio gives insight into the material properties of the silicide layer since the resistance at low temperatures is dominated by impurity and defect scattering~\cite{testardi1977anomalous}.
	For higher annealing temperatures the RRR is greater for thicker layers (with the exception of the \SI{200}{\nano\meter} sample annealed at \SI{750}{\celsius}, which is also an outlier in terms of resistivity), which is consistent with the lower quality of thinner films as reported before~\cite{michikami1982v3si}.
	After annealing at \SI{550}{\celsius}, however, we find that the RRR is highest in samples where only \SI{20}{\nano\meter} of \ce{V3Si} was deposited, followed by \SI{50}{\nano\meter} and \SI{100}{\nano\meter}, in that order.
	This could be an indication that a nearly homogeneous \ce{VSi2} layer has already started to form at the \ce{V3Si}/Si interface in the thinnest of these films, with inhomogeneity increasing for thicker films of 50 and \SI{100}{\nano\meter}.
	Complete homogenization of a \SI{20}{\nano\meter} film during two minutes would suggest however that Si, the dominant diffusing species in \ce{VSi2} formation~\cite{chu1974identification}, has diffused with a diffusivity of at least \SI{1.8}{\nano\meter\squared\per\second}, values that have only been reported for annealing temperatures of \SI{650}{\celsius}~\cite{tu1973formation}.
	The higher RRR at \SI{550}{\celsius} of the sample with \SI{200}{\nano\meter} deposited \ce{V3Si} suggests that this layer is too thick for diffusion to have already caused mixing throughout the entire film, and this relatively high value could be attributed to crystallization of \ce{V3Si}.
	
	The appearance of crystalline \ce{V3Si} at temperatures of around \SI{500}{\celsius} is consistent with the in-situ XRD analysis on non-reactive substrates discussed in section~\ref{sec:xrd} (see especially Fig.~\ref{fig:figiii2}).
	This is further corroborated by the appearance of superconductivity with critical temperatures above \SI{2}{\kelvin} as shown in Fig.~\ref{fig:v3sicombinedplot}c, to be contrasted with the critical temperature of \SI{0.9}{\kelvin} of amorphous \SI{200}{\nano\meter}-thick \ce{V3Si} layers on silicon substrates prepared under identical conditions~\cite{vethaak2021influence}.
	A critical temperature of \SI{10.62}{\kelvin} (10\% and 90\% transitions at 10.32 and \SI{11.00}{\kelvin}, respectively) was attained after annealing at \SI{650}{\celsius} in a \SI{200}{\nano\meter} deposited film, identical to that obtained on samples where \ce{VSi2} formation was prevented by a \SI{20}{\nano\meter} film of \ce{SiO2}, with a superconducting transition that is only 12\% wider~\cite{vethaak2021influence}.
	This suggests that the nearby formation of \ce{VSi2} does not have a significant impact on the superconducting properties of the remaining \ce{V3Si}.
	Meanwhile, the absence of superconductivity above \SI{2}{\kelvin} in samples with 20 or \SI{50}{\nano\meter} deposited \ce{V3Si}, after annealing at any temperature, confirms that Si diffusion and \ce{VSi2} formation rapidly remove \ce{V3Si} or even prevent its crystallization altogether in thinner films.
	
\section{Discussion}
	
	Instead of forming vanadium silicide by reactive diffusion, \ce{V3Si} was directly sputtered from a compound target at the right stoichiometry, giving the advantages of having a sharp interface between the deposited layer and the substrate, and requiring no solid-state reaction to reach the desired phase.
	However, the interface must be better controlled in terms of impurities, and the \ce{V3Si} is amorphous after deposition.
	First, the crystallization was studied on 2 types of substrate, oxidized silicon with a \SI{20}{\nano\meter} thick \ce{SiO2} surface layer and monocrystalline sapphire (\ce{Al2O3}).
	It was clear that no reaction occurred with either of these substrates at any temperature.
	Earlier work showed that the mismatch in TEC with the substrates strongly impacts the stress built up during the growth of \ce{V3Si}, which can lower the superconducting critical temperature~\cite{vethaak2021influence}.
	
	Layer crystallization started at \SI{500}{\celsius}, leading to high tensile stress on either substrate up to \SI{650}{\celsius}.
	This stress then relaxed with increasing temperature, both due to atomic diffusion and the relatively low TEC of the substrates.
	From \SI{650}{\celsius} to around \SI{700}{\celsius}, an increase in tensile stress together with a decrease in the out-of-plane IB can be explained by grain growth in fair agreement with the model of P. Chaudari~\cite{chaudhari1972grain}.
	During cooling, the \ce{V3Si} stress behavior matches perfectly with TEC of silicon and sapphire substrate, indicating a purely thermoelastic behavior.
	It is thus clear that the thermomechanical properties of the substrate strongly impact the stress development in the \ce{V3Si} film, which can be mitigated by proper choice of annealing conditions.
	
	Since \ce{V3Si} is amorphous after deposition, crystallization by subsequent thermal processing is required to achieve the desired superconducting properties.
	However, on a silicon substrate this has the undesired side effect of \ce{VSi2} nucleation.
	In fact, in the thinnest films that were studied, with 20 and \SI{50}{\nano\meter} of deposited silicide, a nearly homogeneous \ce{VSi2} layer had already started to form at the interface with the substrate before any crystallization of \ce{V3Si} was observed by critical temperature measurements.
	Full consumption of the deposited \ce{V3Si} occurred at higher temperatures between 600--\SI{650}{\celsius} and 650--\SI{700}{\celsius} for layers of 100 and \SI{200}{\nano\meter}, respectively, allowing for \ce{V3Si} crystallization, as well as grain growth.
	The critical temperatures thus obtained on a silicon substrate prior to complete \ce{V3Si} consumption are identical, within experimental error, to those obtained on \ce{SiO2}~\cite{vethaak2021influence}, indicating no detrimental effect of diminished thickness or proximity to growing \ce{VSi2} on the quality of the superconductor.
	This is in contrast with earlier results obtained by \ce{V3Si} sputtering from a compound target on heated Si substrates with \SI{800}{\nano\meter} of thermal \ce{SiO2}, where a strong dependence of film thickness was found below \SI{200}{\nano\meter}~\cite{michikami1982v3si}.
	This variation in thickness was at the time attributed to the presence of impurities such as oxygen, while earlier reports had argued that the chemical composition of the layer was unimportant, and any variations that it causes in $T_\text{c}$ are due to a change in lattice parameter~\cite{testardi1970unusual}.
	Since no reaction with the substrate occurs on \ce{SiO2} (see Fig.~\ref{fig:sem_sio2_si}), and the critical temperature is independent of the thickness of crystalline \ce{V3Si} itself, the reduced $T_\text{c}$ reported in this earlier study~\cite{michikami1982v3si} could also be attributed to differences in grain growth, or stress development during cooling from \SI{560}{\celsius}, at which the layers were deposited, to cryogenic temperatures in the range of 10--\SI{15}{\kelvin}.		
	
\section{Conclusion}
	
	Superconducting thin films are of increasing interest as solid-state quantum technologies are scaled up.
	To facilitate large-scale fabrication, as well as co-integration with classical electronics, it is important that the choice of superconducting material is compatible with CMOS technology.
	This report addresses two challenges for the integration of \ce{V3Si}, which as a silicide with a superconducting critical temperature of up to \SI{17}{\kelvin}~\cite{hardy1953superconducting,blumberg1960correlations} is a natural candidate in this context.
	
	The first is the reduction in critical temperature due to thermal strain.
	While \ce{V3Si} is compatible with the oxides of silicon and aluminum, both currently used in VLSI devices, it is necessary to mitigate the stresses induced by the interplay between crystallization, grain growth and TEC mismatch with the substrate.
	It is found that the final stress obtained on either substrate depends strongly on the thermal budget that the samples are subjected to, identifying annealing as a key process step to control the critical temperature of the film.
	
	Second, thermal processing also provides a means of controlling the formation of the undesired phase \ce{VSi2} on HF-cleaned silicon substrates.
	For \ce{V3Si} films with a thickness of \SI{100}{\nano\meter} or more, there exists a process window where it can be crystallized before consumption by the \ce{VSi2} phase, with a temperature range that depends on the thickness of the film.
	
	Direct applications of \ce{V3Si} could be in high-frequency resonators on oxides or oxidized substrates, where the impact of a potential stress residual on the quality factor has to be evaluated.
	A further interest would be to fabricate Josephson junctions, where the weak link could either be an oxide in a vertical geometry, or silicon in a planar junction.
	In the latter case, a Josephson field effect transistor could be fabricated by forming \ce{V3Si} in the source and drain contacts of a CMOS transistor, in which case one is faced with a trade-off between improving the \ce{V3Si} critical temperature and limiting the growth of the \ce{VSi2} phase.

\section{Data availability}
	
	The data that support the findings of this study are available from the corresponding author upon reasonable request.

\section{Acknowledgments}
	
	T.D.V. acknowledges the European Union’s Horizon 2020 research and innovation programme under the Marie Skłodowska-Curie grant agreement No 754303. This work was partially supported by the ANR project SUNISiDEUP (ANR-19-CE47-0010). JX Nippon are gratefully acknowledged for providing the \ce{V3Si} deposition target.

\bibliography{manuscript.bib}

%aipnum4-2.bst 2019-01-14 (MD) hand-edited version of apsrev4-1.bst
\begin{thebibliography}{31}%
\makeatletter
\providecommand \@ifxundefined [1]{%
 \@ifx{#1\undefined}
}%
\providecommand \@ifnum [1]{%
 \ifnum #1\expandafter \@firstoftwo
 \else \expandafter \@secondoftwo
 \fi
}%
\providecommand \@ifx [1]{%
 \ifx #1\expandafter \@firstoftwo
 \else \expandafter \@secondoftwo
 \fi
}%
\providecommand \natexlab [1]{#1}%
\providecommand \enquote  [1]{``#1''}%
\providecommand \bibnamefont  [1]{#1}%
\providecommand \bibfnamefont [1]{#1}%
\providecommand \citenamefont [1]{#1}%
\providecommand \href@noop [0]{\@secondoftwo}%
\providecommand \href [0]{\begingroup \@sanitize@url \@href}%
\providecommand \@href[1]{\@@startlink{#1}\@@href}%
\providecommand \@@href[1]{\endgroup#1\@@endlink}%
\providecommand \@sanitize@url [0]{\catcode `\\12\catcode `\$12\catcode
  `\&12\catcode `\#12\catcode `\^12\catcode `\_12\catcode `\%12\relax}%
\providecommand \@@startlink[1]{}%
\providecommand \@@endlink[0]{}%
\providecommand \url  [0]{\begingroup\@sanitize@url \@url }%
\providecommand \@url [1]{\endgroup\@href {#1}{\urlprefix }}%
\providecommand \urlprefix  [0]{URL }%
\providecommand \Eprint [0]{\href }%
\providecommand \doibase [0]{https://doi.org/}%
\providecommand \selectlanguage [0]{\@gobble}%
\providecommand \bibinfo  [0]{\@secondoftwo}%
\providecommand \bibfield  [0]{\@secondoftwo}%
\providecommand \translation [1]{[#1]}%
\providecommand \BibitemOpen [0]{}%
\providecommand \bibitemStop [0]{}%
\providecommand \bibitemNoStop [0]{.\EOS\space}%
\providecommand \EOS [0]{\spacefactor3000\relax}%
\providecommand \BibitemShut  [1]{\csname bibitem#1\endcsname}%
\let\auto@bib@innerbib\@empty
%</preamble>
\bibitem [{\citenamefont {Arute}\ \emph {et~al.}(2019)\citenamefont {Arute},
  \citenamefont {Arya}, \citenamefont {Babbush}, \citenamefont {Bacon},
  \citenamefont {Bardin}, \citenamefont {Barends}, \citenamefont {Biswas},
  \citenamefont {Boixo}, \citenamefont {Brandao}, \citenamefont {Buell} \emph
  {et~al.}}]{arute2019quantum}%
  \BibitemOpen
  \bibfield  {author} {\bibinfo {author} {\bibfnamefont {F.}~\bibnamefont
  {Arute}}, \bibinfo {author} {\bibfnamefont {K.}~\bibnamefont {Arya}},
  \bibinfo {author} {\bibfnamefont {R.}~\bibnamefont {Babbush}}, \bibinfo
  {author} {\bibfnamefont {D.}~\bibnamefont {Bacon}}, \bibinfo {author}
  {\bibfnamefont {J.~C.}\ \bibnamefont {Bardin}}, \bibinfo {author}
  {\bibfnamefont {R.}~\bibnamefont {Barends}}, \bibinfo {author} {\bibfnamefont
  {R.}~\bibnamefont {Biswas}}, \bibinfo {author} {\bibfnamefont
  {S.}~\bibnamefont {Boixo}}, \bibinfo {author} {\bibfnamefont {F.~G.}\
  \bibnamefont {Brandao}}, \bibinfo {author} {\bibfnamefont {D.~A.}\
  \bibnamefont {Buell}}, \emph {et~al.},\ }\bibfield  {title} {\enquote
  {\bibinfo {title} {Quantum supremacy using a programmable superconducting
  processor},}\ }\href@noop {} {\bibfield  {journal} {\bibinfo  {journal}
  {Nature}\ }\textbf {\bibinfo {volume} {574}},\ \bibinfo {pages} {505--510}
  (\bibinfo {year} {2019})}\BibitemShut {NoStop}%
\bibitem [{\citenamefont {Zwerver}\ \emph {et~al.}(2021)\citenamefont
  {Zwerver}, \citenamefont {Kr{\"a}henmann}, \citenamefont {Watson},
  \citenamefont {Lampert}, \citenamefont {George}, \citenamefont
  {Pillarisetty}, \citenamefont {Bojarski}, \citenamefont {Amin}, \citenamefont
  {Amitonov}, \citenamefont {Boter} \emph {et~al.}}]{zwerver2021qubits}%
  \BibitemOpen
  \bibfield  {author} {\bibinfo {author} {\bibfnamefont {A.}~\bibnamefont
  {Zwerver}}, \bibinfo {author} {\bibfnamefont {T.}~\bibnamefont
  {Kr{\"a}henmann}}, \bibinfo {author} {\bibfnamefont {T.}~\bibnamefont
  {Watson}}, \bibinfo {author} {\bibfnamefont {L.}~\bibnamefont {Lampert}},
  \bibinfo {author} {\bibfnamefont {H.}~\bibnamefont {George}}, \bibinfo
  {author} {\bibfnamefont {R.}~\bibnamefont {Pillarisetty}}, \bibinfo {author}
  {\bibfnamefont {S.}~\bibnamefont {Bojarski}}, \bibinfo {author}
  {\bibfnamefont {P.}~\bibnamefont {Amin}}, \bibinfo {author} {\bibfnamefont
  {S.}~\bibnamefont {Amitonov}}, \bibinfo {author} {\bibfnamefont
  {J.}~\bibnamefont {Boter}}, \emph {et~al.},\ }\bibfield  {title} {\enquote
  {\bibinfo {title} {Qubits made by advanced semiconductor manufacturing},}\
  }\href@noop {} {\bibfield  {journal} {\bibinfo  {journal} {arXiv preprint
  arXiv:2101.12650}\ } (\bibinfo {year} {2021})}\BibitemShut {NoStop}%
\bibitem [{\citenamefont {Majer}\ \emph {et~al.}(2007)\citenamefont {Majer},
  \citenamefont {Chow}, \citenamefont {Gambetta}, \citenamefont {Koch},
  \citenamefont {Johnson}, \citenamefont {Schreier}, \citenamefont {Frunzio},
  \citenamefont {Schuster}, \citenamefont {Houck}, \citenamefont {Wallraff}
  \emph {et~al.}}]{majer2007coupling}%
  \BibitemOpen
  \bibfield  {author} {\bibinfo {author} {\bibfnamefont {J.}~\bibnamefont
  {Majer}}, \bibinfo {author} {\bibfnamefont {J.}~\bibnamefont {Chow}},
  \bibinfo {author} {\bibfnamefont {J.}~\bibnamefont {Gambetta}}, \bibinfo
  {author} {\bibfnamefont {J.}~\bibnamefont {Koch}}, \bibinfo {author}
  {\bibfnamefont {B.}~\bibnamefont {Johnson}}, \bibinfo {author} {\bibfnamefont
  {J.}~\bibnamefont {Schreier}}, \bibinfo {author} {\bibfnamefont
  {L.}~\bibnamefont {Frunzio}}, \bibinfo {author} {\bibfnamefont
  {D.}~\bibnamefont {Schuster}}, \bibinfo {author} {\bibfnamefont {A.~A.}\
  \bibnamefont {Houck}}, \bibinfo {author} {\bibfnamefont {A.}~\bibnamefont
  {Wallraff}}, \emph {et~al.},\ }\bibfield  {title} {\enquote {\bibinfo {title}
  {Coupling superconducting qubits via a cavity bus},}\ }\href@noop {}
  {\bibfield  {journal} {\bibinfo  {journal} {Nature}\ }\textbf {\bibinfo
  {volume} {449}},\ \bibinfo {pages} {443--447} (\bibinfo {year}
  {2007})}\BibitemShut {NoStop}%
\bibitem [{\citenamefont {Mi}\ \emph {et~al.}(2018)\citenamefont {Mi},
  \citenamefont {Benito}, \citenamefont {Putz}, \citenamefont {Zajac},
  \citenamefont {Taylor}, \citenamefont {Burkard},\ and\ \citenamefont
  {Petta}}]{mi2018coherent}%
  \BibitemOpen
  \bibfield  {author} {\bibinfo {author} {\bibfnamefont {X.}~\bibnamefont
  {Mi}}, \bibinfo {author} {\bibfnamefont {M.}~\bibnamefont {Benito}}, \bibinfo
  {author} {\bibfnamefont {S.}~\bibnamefont {Putz}}, \bibinfo {author}
  {\bibfnamefont {D.~M.}\ \bibnamefont {Zajac}}, \bibinfo {author}
  {\bibfnamefont {J.~M.}\ \bibnamefont {Taylor}}, \bibinfo {author}
  {\bibfnamefont {G.}~\bibnamefont {Burkard}},\ and\ \bibinfo {author}
  {\bibfnamefont {J.~R.}\ \bibnamefont {Petta}},\ }\bibfield  {title} {\enquote
  {\bibinfo {title} {A coherent spin--photon interface in silicon},}\
  }\href@noop {} {\bibfield  {journal} {\bibinfo  {journal} {Nature}\ }\textbf
  {\bibinfo {volume} {555}},\ \bibinfo {pages} {599--603} (\bibinfo {year}
  {2018})}\BibitemShut {NoStop}%
\bibitem [{\citenamefont {Shibata}\ \emph {et~al.}(1981)\citenamefont
  {Shibata}, \citenamefont {Hieda}, \citenamefont {Sato}, \citenamefont
  {Konaka}, \citenamefont {Dang},\ and\ \citenamefont
  {Iizuka}}]{shibata1981optimally}%
  \BibitemOpen
  \bibfield  {author} {\bibinfo {author} {\bibfnamefont {T.}~\bibnamefont
  {Shibata}}, \bibinfo {author} {\bibfnamefont {K.}~\bibnamefont {Hieda}},
  \bibinfo {author} {\bibfnamefont {M.}~\bibnamefont {Sato}}, \bibinfo {author}
  {\bibfnamefont {M.}~\bibnamefont {Konaka}}, \bibinfo {author} {\bibfnamefont
  {R.}~\bibnamefont {Dang}},\ and\ \bibinfo {author} {\bibfnamefont
  {H.}~\bibnamefont {Iizuka}},\ }\bibfield  {title} {\enquote {\bibinfo {title}
  {An optimally designed process for submicron mosfets},}\ }in\ \href@noop {}
  {\emph {\bibinfo {booktitle} {1981 International Electron Devices Meeting}}}\
  (\bibinfo {organization} {IEEE},\ \bibinfo {year} {1981})\ pp.\ \bibinfo
  {pages} {647--650}\BibitemShut {NoStop}%
\bibitem [{\citenamefont {Zhang}\ and\ \citenamefont
  {Zhang}(2014)}]{zhang2014metal}%
  \BibitemOpen
  \bibfield  {author} {\bibinfo {author} {\bibfnamefont {S.-L.}\ \bibnamefont
  {Zhang}}\ and\ \bibinfo {author} {\bibfnamefont {Z.}~\bibnamefont {Zhang}},\
  }\bibfield  {title} {\enquote {\bibinfo {title} {Metal silicides in advanced
  complementary metal-oxide-semiconductor (cmos) technology},}\ }in\ \href@noop
  {} {\emph {\bibinfo {booktitle} {Metallic Films for Electronic, Optical and
  Magnetic Applications}}},\ \bibinfo {editor} {edited by\ \bibinfo {editor}
  {\bibfnamefont {K.}~\bibnamefont {Barmak}}\ and\ \bibinfo {editor}
  {\bibfnamefont {K.}~\bibnamefont {Coffey}}}\ (\bibinfo  {publisher} {Woodhead
  Publishing},\ \bibinfo {year} {2014})\ pp.\ \bibinfo {pages}
  {244--301}\BibitemShut {NoStop}%
\bibitem [{\citenamefont {Hardy}\ and\ \citenamefont
  {Hulm}(1953)}]{hardy1953superconducting}%
  \BibitemOpen
  \bibfield  {author} {\bibinfo {author} {\bibfnamefont {G.~F.}\ \bibnamefont
  {Hardy}}\ and\ \bibinfo {author} {\bibfnamefont {J.~K.}\ \bibnamefont
  {Hulm}},\ }\bibfield  {title} {\enquote {\bibinfo {title} {Superconducting
  silicides and germanides},}\ }\href@noop {} {\bibfield  {journal} {\bibinfo
  {journal} {Physical Review}\ }\textbf {\bibinfo {volume} {89}},\ \bibinfo
  {pages} {884} (\bibinfo {year} {1953})}\BibitemShut {NoStop}%
\bibitem [{\citenamefont {Blumberg}\ \emph {et~al.}(1960)\citenamefont
  {Blumberg}, \citenamefont {Eisinger}, \citenamefont {Jaccarino},\ and\
  \citenamefont {Matthias}}]{blumberg1960correlations}%
  \BibitemOpen
  \bibfield  {author} {\bibinfo {author} {\bibfnamefont {W.}~\bibnamefont
  {Blumberg}}, \bibinfo {author} {\bibfnamefont {J.}~\bibnamefont {Eisinger}},
  \bibinfo {author} {\bibfnamefont {V.}~\bibnamefont {Jaccarino}},\ and\
  \bibinfo {author} {\bibfnamefont {B.}~\bibnamefont {Matthias}},\ }\bibfield
  {title} {\enquote {\bibinfo {title} {Correlations between superconductivity
  and nuclear magnetic resonance properties},}\ }\href@noop {} {\bibfield
  {journal} {\bibinfo  {journal} {Physical Review Letters}\ }\textbf {\bibinfo
  {volume} {5}},\ \bibinfo {pages} {149} (\bibinfo {year} {1960})}\BibitemShut
  {NoStop}%
\bibitem [{\citenamefont {Vethaak}\ \emph {et~al.}(2021)\citenamefont
  {Vethaak}, \citenamefont {Gustavo}, \citenamefont {Farjot}, \citenamefont
  {Kubart}, \citenamefont {Gergaud}, \citenamefont {Zhang}, \citenamefont
  {Nemouchi},\ and\ \citenamefont {Lefloch}}]{vethaak2021influence}%
  \BibitemOpen
  \bibfield  {author} {\bibinfo {author} {\bibfnamefont {T.~D.}\ \bibnamefont
  {Vethaak}}, \bibinfo {author} {\bibfnamefont {F.}~\bibnamefont {Gustavo}},
  \bibinfo {author} {\bibfnamefont {T.}~\bibnamefont {Farjot}}, \bibinfo
  {author} {\bibfnamefont {T.}~\bibnamefont {Kubart}}, \bibinfo {author}
  {\bibfnamefont {P.}~\bibnamefont {Gergaud}}, \bibinfo {author} {\bibfnamefont
  {S.-L.}\ \bibnamefont {Zhang}}, \bibinfo {author} {\bibfnamefont
  {F.}~\bibnamefont {Nemouchi}},\ and\ \bibinfo {author} {\bibfnamefont
  {F.}~\bibnamefont {Lefloch}},\ }\bibfield  {title} {\enquote {\bibinfo
  {title} {Influence of substrate-induced thermal stress on the superconducting
  properties of v 3 si thin films},}\ }\href@noop {} {\bibfield  {journal}
  {\bibinfo  {journal} {Journal of Applied Physics}\ }\textbf {\bibinfo
  {volume} {129}},\ \bibinfo {pages} {105104} (\bibinfo {year}
  {2021})}\BibitemShut {NoStop}%
\bibitem [{\citenamefont {Kr{\"a}utle}, \citenamefont {Nicolet},\ and\
  \citenamefont {Mayer}(1974)}]{krautle1974kinetics}%
  \BibitemOpen
  \bibfield  {author} {\bibinfo {author} {\bibfnamefont {H.}~\bibnamefont
  {Kr{\"a}utle}}, \bibinfo {author} {\bibfnamefont {M.-A.}\ \bibnamefont
  {Nicolet}},\ and\ \bibinfo {author} {\bibfnamefont {J.}~\bibnamefont
  {Mayer}},\ }\bibfield  {title} {\enquote {\bibinfo {title} {Kinetics of
  silicide formation by thin films of v on si and \ce{SiO2} substrates},}\
  }\href@noop {} {\bibfield  {journal} {\bibinfo  {journal} {Journal of Applied
  Physics}\ }\textbf {\bibinfo {volume} {45}},\ \bibinfo {pages} {3304--3308}
  (\bibinfo {year} {1974})}\BibitemShut {NoStop}%
\bibitem [{\citenamefont {Pretorius}, \citenamefont {Marais},\ and\
  \citenamefont {Theron}(1993)}]{pretorius1993thin}%
  \BibitemOpen
  \bibfield  {author} {\bibinfo {author} {\bibfnamefont {R.}~\bibnamefont
  {Pretorius}}, \bibinfo {author} {\bibfnamefont {T.}~\bibnamefont {Marais}},\
  and\ \bibinfo {author} {\bibfnamefont {C.}~\bibnamefont {Theron}},\
  }\bibfield  {title} {\enquote {\bibinfo {title} {Thin film compound phase
  formation sequence: An effective heat of formation model},}\ }\href@noop {}
  {\bibfield  {journal} {\bibinfo  {journal} {Materials Science Reports}\
  }\textbf {\bibinfo {volume} {10}},\ \bibinfo {pages} {1--83} (\bibinfo {year}
  {1993})}\BibitemShut {NoStop}%
\bibitem [{\citenamefont {Michikami}\ and\ \citenamefont
  {Takenaka}(1982)}]{michikami1982v3si}%
  \BibitemOpen
  \bibfield  {author} {\bibinfo {author} {\bibfnamefont {O.}~\bibnamefont
  {Michikami}}\ and\ \bibinfo {author} {\bibfnamefont {H.}~\bibnamefont
  {Takenaka}},\ }\bibfield  {title} {\enquote {\bibinfo {title} {\ce{V3Si}
  thin-film synthesis by magnetron sputtering},}\ }\href@noop {} {\bibfield
  {journal} {\bibinfo  {journal} {Japanese Journal of Applied Physics}\
  }\textbf {\bibinfo {volume} {21}},\ \bibinfo {pages} {L149} (\bibinfo {year}
  {1982})}\BibitemShut {NoStop}%
\bibitem [{\citenamefont {Theuerer}\ and\ \citenamefont
  {Hauser}(1964)}]{theuerer1964getter}%
  \BibitemOpen
  \bibfield  {author} {\bibinfo {author} {\bibfnamefont {H.~C.}\ \bibnamefont
  {Theuerer}}\ and\ \bibinfo {author} {\bibfnamefont {J.}~\bibnamefont
  {Hauser}},\ }\bibfield  {title} {\enquote {\bibinfo {title} {Getter
  sputtering for the preparation of thin films of superconducting elements and
  compounds},}\ }\href@noop {} {\bibfield  {journal} {\bibinfo  {journal}
  {Journal of Applied Physics}\ }\textbf {\bibinfo {volume} {35}},\ \bibinfo
  {pages} {554--555} (\bibinfo {year} {1964})}\BibitemShut {NoStop}%
\bibitem [{\citenamefont {Testardi}\ \emph {et~al.}(1970)\citenamefont
  {Testardi}, \citenamefont {Kunzler}, \citenamefont {Levinstein},\ and\
  \citenamefont {Wernick}}]{testardi1970unusual}%
  \BibitemOpen
  \bibfield  {author} {\bibinfo {author} {\bibfnamefont {L.}~\bibnamefont
  {Testardi}}, \bibinfo {author} {\bibfnamefont {J.}~\bibnamefont {Kunzler}},
  \bibinfo {author} {\bibfnamefont {H.}~\bibnamefont {Levinstein}},\ and\
  \bibinfo {author} {\bibfnamefont {J.}~\bibnamefont {Wernick}},\ }\bibfield
  {title} {\enquote {\bibinfo {title} {Unusual strain dependence of tc and
  related effects in a-15 superconductors},}\ }\href@noop {} {\bibfield
  {journal} {\bibinfo  {journal} {Solid State Communications}\ }\textbf
  {\bibinfo {volume} {8}},\ \bibinfo {pages} {907--911} (\bibinfo {year}
  {1970})}\BibitemShut {NoStop}%
\bibitem [{\citenamefont {Testardi}(1971)}]{testardi1971unusual95}%
  \BibitemOpen
  \bibfield  {author} {\bibinfo {author} {\bibfnamefont {L.}~\bibnamefont
  {Testardi}},\ }\bibfield  {title} {\enquote {\bibinfo {title} {Unusual strain
  dependence of t c and related effects for high-temperature (a- 15-structure)
  superconductors: Sound velocity at the superconducting phase transition},}\
  }\href@noop {} {\bibfield  {journal} {\bibinfo  {journal} {Physical Review
  B}\ }\textbf {\bibinfo {volume} {3}},\ \bibinfo {pages} {95} (\bibinfo {year}
  {1971})}\BibitemShut {NoStop}%
\bibitem [{\citenamefont {Testardi}\ \emph {et~al.}(1971)\citenamefont
  {Testardi}, \citenamefont {Kunzler}, \citenamefont {Levinstein},
  \citenamefont {Maita},\ and\ \citenamefont {Wernick}}]{testardi1971unusual}%
  \BibitemOpen
  \bibfield  {author} {\bibinfo {author} {\bibfnamefont {L.}~\bibnamefont
  {Testardi}}, \bibinfo {author} {\bibfnamefont {J.}~\bibnamefont {Kunzler}},
  \bibinfo {author} {\bibfnamefont {H.}~\bibnamefont {Levinstein}}, \bibinfo
  {author} {\bibfnamefont {J.}~\bibnamefont {Maita}},\ and\ \bibinfo {author}
  {\bibfnamefont {J.}~\bibnamefont {Wernick}},\ }\bibfield  {title} {\enquote
  {\bibinfo {title} {Unusual strain dependence of tc and related effects for
  high-temperature (a-15-structure) superconductors: Elastic, thermal, and
  alloy behavior},}\ }\href@noop {} {\bibfield  {journal} {\bibinfo  {journal}
  {Physical Review B}\ }\textbf {\bibinfo {volume} {3}},\ \bibinfo {pages}
  {107} (\bibinfo {year} {1971})}\BibitemShut {NoStop}%
\bibitem [{\citenamefont {Testardi}(1972)}]{testardi1972structural}%
  \BibitemOpen
  \bibfield  {author} {\bibinfo {author} {\bibfnamefont {L.}~\bibnamefont
  {Testardi}},\ }\bibfield  {title} {\enquote {\bibinfo {title} {Structural
  instability, anharmonicity, and high-temperature superconductivity in a-
  15-structure compounds},}\ }\href@noop {} {\bibfield  {journal} {\bibinfo
  {journal} {Physical Review B}\ }\textbf {\bibinfo {volume} {5}},\ \bibinfo
  {pages} {4342} (\bibinfo {year} {1972})}\BibitemShut {NoStop}%
\bibitem [{\citenamefont {Batterman}\ and\ \citenamefont
  {Barrett}(1964)}]{batterman1964crystal}%
  \BibitemOpen
  \bibfield  {author} {\bibinfo {author} {\bibfnamefont {B.}~\bibnamefont
  {Batterman}}\ and\ \bibinfo {author} {\bibfnamefont {C.}~\bibnamefont
  {Barrett}},\ }\bibfield  {title} {\enquote {\bibinfo {title} {Crystal
  structure of superconducting \ce{V3Si}},}\ }\href@noop {} {\bibfield
  {journal} {\bibinfo  {journal} {Physical Review Letters}\ }\textbf {\bibinfo
  {volume} {13}},\ \bibinfo {pages} {390} (\bibinfo {year} {1964})}\BibitemShut
  {NoStop}%
\bibitem [{\citenamefont {Batterman}\ and\ \citenamefont
  {Barrett}(1966)}]{batterman1966low}%
  \BibitemOpen
  \bibfield  {author} {\bibinfo {author} {\bibfnamefont {B.}~\bibnamefont
  {Batterman}}\ and\ \bibinfo {author} {\bibfnamefont {C.}~\bibnamefont
  {Barrett}},\ }\bibfield  {title} {\enquote {\bibinfo {title} {Low-temperature
  structural transformation in \ce{V3Si}},}\ }\href@noop {} {\bibfield
  {journal} {\bibinfo  {journal} {Physical Review}\ }\textbf {\bibinfo {volume}
  {145}},\ \bibinfo {pages} {296} (\bibinfo {year} {1966})}\BibitemShut
  {NoStop}%
\bibitem [{\citenamefont {Noyan}\ and\ \citenamefont
  {Cohen}(2013)}]{noyan2013residual}%
  \BibitemOpen
  \bibfield  {author} {\bibinfo {author} {\bibfnamefont {I.~C.}\ \bibnamefont
  {Noyan}}\ and\ \bibinfo {author} {\bibfnamefont {J.~B.}\ \bibnamefont
  {Cohen}},\ }\href@noop {} {\emph {\bibinfo {title} {Residual stress:
  measurement by diffraction and interpretation}}}\ (\bibinfo  {publisher}
  {Springer},\ \bibinfo {year} {2013})\BibitemShut {NoStop}%
\bibitem [{\citenamefont {Murray}(2013)}]{murray2013equivalence}%
  \BibitemOpen
  \bibfield  {author} {\bibinfo {author} {\bibfnamefont {C.~E.}\ \bibnamefont
  {Murray}},\ }\bibfield  {title} {\enquote {\bibinfo {title} {Equivalence of
  kr{\"o}ner and weighted voigt-reuss models for x-ray stress determination},}\
  }\href@noop {} {\bibfield  {journal} {\bibinfo  {journal} {Journal of Applied
  Physics}\ }\textbf {\bibinfo {volume} {113}},\ \bibinfo {pages} {153509}
  (\bibinfo {year} {2013})}\BibitemShut {NoStop}%
\bibitem [{\citenamefont {Gaillac}, \citenamefont {Pullumbi},\ and\
  \citenamefont {Coudert}(2016)}]{gaillac2016elate}%
  \BibitemOpen
  \bibfield  {author} {\bibinfo {author} {\bibfnamefont {R.}~\bibnamefont
  {Gaillac}}, \bibinfo {author} {\bibfnamefont {P.}~\bibnamefont {Pullumbi}},\
  and\ \bibinfo {author} {\bibfnamefont {F.-X.}\ \bibnamefont {Coudert}},\
  }\bibfield  {title} {\enquote {\bibinfo {title} {Elate: an open-source online
  application for analysis and visualization of elastic tensors},}\ }\href@noop
  {} {\bibfield  {journal} {\bibinfo  {journal} {Journal of Physics: Condensed
  Matter}\ }\textbf {\bibinfo {volume} {28}},\ \bibinfo {pages} {275201}
  (\bibinfo {year} {2016})},\ \bibinfo {note}
  {\url{http://progs.coudert.name/elate/mp?query=mp-2567}}\BibitemShut
  {NoStop}%
\bibitem [{\citenamefont {Scherrer}(1918)}]{scherrer1918nachr}%
  \BibitemOpen
  \bibfield  {author} {\bibinfo {author} {\bibfnamefont {P.}~\bibnamefont
  {Scherrer}},\ }\bibfield  {title} {\enquote {\bibinfo {title} {Nachrichten
  von der gesellschaft der wissenschaften zu g{\"o}ttingen},}\ }\href@noop {}
  {\bibfield  {journal} {\bibinfo  {journal} {Mathematisch-Physikalische
  Klasse}\ }\textbf {\bibinfo {volume} {2}},\ \bibinfo {pages} {98--100}
  (\bibinfo {year} {1918})}\BibitemShut {NoStop}%
\bibitem [{\citenamefont {Cullity}\ and\ \citenamefont
  {Stock}(2013)}]{cullity2001elements}%
  \BibitemOpen
  \bibfield  {author} {\bibinfo {author} {\bibfnamefont {B.}~\bibnamefont
  {Cullity}}\ and\ \bibinfo {author} {\bibfnamefont {S.}~\bibnamefont
  {Stock}},\ }\href@noop {} {\emph {\bibinfo {title} {Elements of x-ray
  diffraction}}}\ (\bibinfo  {publisher} {Pearson Education Limited},\ \bibinfo
  {year} {2013})\BibitemShut {NoStop}%
\bibitem [{\citenamefont {Vodenitcharova}\ \emph {et~al.}(2006)\citenamefont
  {Vodenitcharova}, \citenamefont {Zhang}, \citenamefont {Zarudi},
  \citenamefont {Yin}, \citenamefont {Domyo},\ and\ \citenamefont
  {Ho}}]{vodenitcharova2006effect}%
  \BibitemOpen
  \bibfield  {author} {\bibinfo {author} {\bibfnamefont {T.}~\bibnamefont
  {Vodenitcharova}}, \bibinfo {author} {\bibfnamefont {L.}~\bibnamefont
  {Zhang}}, \bibinfo {author} {\bibfnamefont {I.}~\bibnamefont {Zarudi}},
  \bibinfo {author} {\bibfnamefont {Y.}~\bibnamefont {Yin}}, \bibinfo {author}
  {\bibfnamefont {H.}~\bibnamefont {Domyo}},\ and\ \bibinfo {author}
  {\bibfnamefont {T.}~\bibnamefont {Ho}},\ }\bibfield  {title} {\enquote
  {\bibinfo {title} {The effect of thermal shocks on the stresses in a sapphire
  wafer},}\ }\href@noop {} {\bibfield  {journal} {\bibinfo  {journal} {IEEE
  transactions on semiconductor manufacturing}\ }\textbf {\bibinfo {volume}
  {19}},\ \bibinfo {pages} {449--454} (\bibinfo {year} {2006})}\BibitemShut
  {NoStop}%
\bibitem [{\citenamefont {Watanabe}, \citenamefont {Yamada},\ and\
  \citenamefont {Okaji}(2004)}]{watanabe2004linear}%
  \BibitemOpen
  \bibfield  {author} {\bibinfo {author} {\bibfnamefont {H.}~\bibnamefont
  {Watanabe}}, \bibinfo {author} {\bibfnamefont {N.}~\bibnamefont {Yamada}},\
  and\ \bibinfo {author} {\bibfnamefont {M.}~\bibnamefont {Okaji}},\ }\bibfield
   {title} {\enquote {\bibinfo {title} {Linear thermal expansion coefficient of
  silicon from 293 to 1000 k},}\ }\href@noop {} {\bibfield  {journal} {\bibinfo
   {journal} {International journal of thermophysics}\ }\textbf {\bibinfo
  {volume} {25}},\ \bibinfo {pages} {221--236} (\bibinfo {year}
  {2004})}\BibitemShut {NoStop}%
\bibitem [{\citenamefont {Chu}\ \emph {et~al.}(1974)\citenamefont {Chu},
  \citenamefont {Kra{\"u}tle}, \citenamefont {Mayer}, \citenamefont
  {M{\"u}ller}, \citenamefont {Nicolet},\ and\ \citenamefont
  {Tu}}]{chu1974identification}%
  \BibitemOpen
  \bibfield  {author} {\bibinfo {author} {\bibfnamefont {W.}~\bibnamefont
  {Chu}}, \bibinfo {author} {\bibfnamefont {H.}~\bibnamefont {Kra{\"u}tle}},
  \bibinfo {author} {\bibfnamefont {J.}~\bibnamefont {Mayer}}, \bibinfo
  {author} {\bibfnamefont {H.}~\bibnamefont {M{\"u}ller}}, \bibinfo {author}
  {\bibfnamefont {M.-A.}\ \bibnamefont {Nicolet}},\ and\ \bibinfo {author}
  {\bibfnamefont {K.}~\bibnamefont {Tu}},\ }\bibfield  {title} {\enquote
  {\bibinfo {title} {Identification of the dominant diffusing species in
  silicide formation},}\ }\href@noop {} {\bibfield  {journal} {\bibinfo
  {journal} {Applied Physics Letters}\ }\textbf {\bibinfo {volume} {25}},\
  \bibinfo {pages} {454--457} (\bibinfo {year} {1974})}\BibitemShut {NoStop}%
\bibitem [{\citenamefont {Schutz}\ and\ \citenamefont
  {Testardi}(1979)}]{schutz1979formation}%
  \BibitemOpen
  \bibfield  {author} {\bibinfo {author} {\bibfnamefont {R.}~\bibnamefont
  {Schutz}}\ and\ \bibinfo {author} {\bibfnamefont {L.}~\bibnamefont
  {Testardi}},\ }\bibfield  {title} {\enquote {\bibinfo {title} {The formation
  of vanadium silicides at thin-film interfaces},}\ }\href@noop {} {\bibfield
  {journal} {\bibinfo  {journal} {Journal of Applied Physics}\ }\textbf
  {\bibinfo {volume} {50}},\ \bibinfo {pages} {5773--5781} (\bibinfo {year}
  {1979})}\BibitemShut {NoStop}%
\bibitem [{\citenamefont {Testardi}, \citenamefont {Poate},\ and\ \citenamefont
  {Levinstein}(1977)}]{testardi1977anomalous}%
  \BibitemOpen
  \bibfield  {author} {\bibinfo {author} {\bibfnamefont {L.}~\bibnamefont
  {Testardi}}, \bibinfo {author} {\bibfnamefont {J.}~\bibnamefont {Poate}},\
  and\ \bibinfo {author} {\bibfnamefont {H.}~\bibnamefont {Levinstein}},\
  }\bibfield  {title} {\enquote {\bibinfo {title} {Anomalous electrical
  resistivity and defects in a-15 compounds},}\ }\href@noop {} {\bibfield
  {journal} {\bibinfo  {journal} {Physical Review B}\ }\textbf {\bibinfo
  {volume} {15}},\ \bibinfo {pages} {2570} (\bibinfo {year}
  {1977})}\BibitemShut {NoStop}%
\bibitem [{\citenamefont {Tu}, \citenamefont {Ziegler},\ and\ \citenamefont
  {Kircher}(1973)}]{tu1973formation}%
  \BibitemOpen
  \bibfield  {author} {\bibinfo {author} {\bibfnamefont {K.}~\bibnamefont
  {Tu}}, \bibinfo {author} {\bibfnamefont {J.}~\bibnamefont {Ziegler}},\ and\
  \bibinfo {author} {\bibfnamefont {C.}~\bibnamefont {Kircher}},\ }\bibfield
  {title} {\enquote {\bibinfo {title} {Formation of vanadium silicides by the
  interactions of v with bare and oxidized si wafers},}\ }\href@noop {}
  {\bibfield  {journal} {\bibinfo  {journal} {Applied Physics Letters}\
  }\textbf {\bibinfo {volume} {23}},\ \bibinfo {pages} {493--495} (\bibinfo
  {year} {1973})}\BibitemShut {NoStop}%
\bibitem [{\citenamefont {Chaudhari}(1972)}]{chaudhari1972grain}%
  \BibitemOpen
  \bibfield  {author} {\bibinfo {author} {\bibfnamefont {P.}~\bibnamefont
  {Chaudhari}},\ }\bibfield  {title} {\enquote {\bibinfo {title} {Grain growth
  and stress relief in thin films},}\ }\href@noop {} {\bibfield  {journal}
  {\bibinfo  {journal} {Journal of Vacuum Science and technology}\ }\textbf
  {\bibinfo {volume} {9}},\ \bibinfo {pages} {520--522} (\bibinfo {year}
  {1972})}\BibitemShut {NoStop}%
\end{thebibliography}%

\end{document}